\documentstyle[11pt,newpasp,twoside,epsf]{article} 
\def\edcomment#1{\iffalse\marginpar{\raggedright\sl#1\/}\else\relax\fi}
\reversemarginpar

\begin{document} \title{Ellipsoidal variability of symbiotic giants} 

\author{J. Miko{\l}ajewska}
\affil{N. Copernicus Astronomical Center, Warsaw, Poland}

\author{E.A. Kolotilov, S. Yu. Shugarov, A.A. Tatarnikova, B.F. Yudin} 
\affil{Sternberg Astronomical Institute, Moscow, Russia}

\begin{abstract} One of fundamental questions in relation to symbiotic binaries is the 
process of mass transfer -- Roche lobe overflow or stellar wind -- and possibility of an 
accretion disc formation. The presence of secondary minima in the quiescent light curves 
of YY Her, CI Cyg and BF Cyg apparently due to ellipsoidal changes of the red giant 
provides here important clues. \end{abstract}

\section{Introduction}

Amongst the evidence for the predominance of wind-accretion is the fact that ellipsoidal 
variations, characteristic of tidally distorted stars are rarely observed for symbiotic 
binaries. However, the general absence of the tidally 
distorted giants in the symbiotic binaries  can be in fact due to the lack of systematic 
searches for the ellipsoidal variations in the red and near-IR range where the cool giant 
dominate the continuum light. This can be best illustrated by the eclipsing symbiotic 
systems  BF Cyg (Fig. 1) and CI Cyg, (Fig. 2; Miko{\l}ajewska 2001).  In both systems, 
during the hot component outburst and its decline, eclipses in the $UBV$ bands are narrow  
with well-defined eclipse contacts whereas at quiescence very broad minima and continuous 
nearly sinusoidal variation are observed. In addition, the quiescent  $VRI$ and near-IR 
light curves show a modulation with half-orbital period as expected for ellipsoidal 
variability of the red giant. 

\section{Analysis of the quiescent light curves of YY Her}

\begin{figure}[t] \plotone{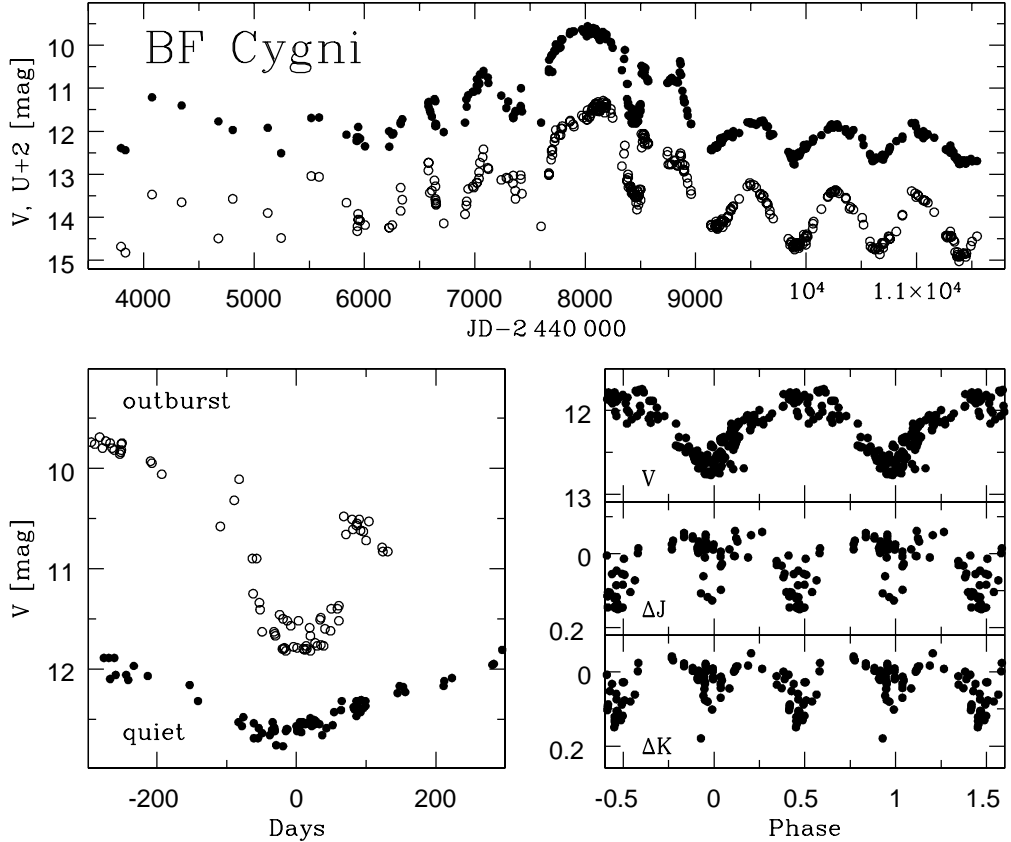} \caption{Optical and near infrared light curves 
of BF~Cyg. As in CI~Cyg (see Fig. 2 of Miko{\l}ajewska 2001),  during the outburst the 
eclipse was narrow with well-defined eclipse contacts whereas the quiescent light curves 
show very broad minima and almost continuous sinusoidal variation (left). The ellipsoidal 
variability of the red giant is visible only in quiescent near-IR light curves (right).} 
\end{figure} 

Similar behaviour has been recently found in YY Her. In particular, the 
photometric and spectroscopic study of YY Her during  the major outburst in 1993  and its 
decline (Tatarnikova et al. 2001) has shown that the minima are due to obscuration of the 
hot component and the ionized nebula. The return to quiescence in 1997/98 was accompanied 
by significant changes in the shape of light curves, and in particular by the appearance 
of a secondary minimum in the $VRI$ light curves (Fig. 2). Tatarnikova et al. (2001)  
interpreted the secondary minimum in terms of ellipsoidal changes of the red giant, 
whereas Hric et al. (2001, hereafter H01) argued that the secondary minimum is caused by 
an eclipse of the red giant by a circumstellar envelope around the hot component. 

The salient characteristic of the quiescent light curves, namely: (i) nearly 
sinusoidal shape of the $U$ light curves; (ii) a flat toped $B$ light curve; and (iii) the 
two minima of almost comparable depth in $I'$ light, combined with the fact that the 
quiescent, ultraviolet + optical,  spectra of YY Her can be satisfactorily matched with a 
three-component model composed of a cool giant dominating the red spectral range, a 
gaseous nebula dominating the near UV range, and the hot component predominant in the 
short UV flux, led us to a simple phenomenological model in 
which the light changes are described by a combination of variable nebular emission and 
ellipsoidal changes of the red giant (Miko{\l}ajewska et al. 2002a, hereafter 
M02).

Each light curve has been converted to fluxes, $F_\lambda = 10^{-0.4 m_{\lambda}}$, and 
then independently fitted by the Fourier cosine series \begin{equation} F_{\lambda} = 
F_{\lambda,0} - A_{\rm n,\lambda} \cos \phi - A_{\rm g,\lambda} \cos 2 \phi. 
\end{equation} In the case of $U$ data, the best fit is obtained with the $\cos \phi$ term 
only. This indicates that the contribution of the giant to the $U$ light is negligible, 
and that all changes can be attributed to the nebular emission. Assuming that the 
amplitude of the nebular emission changes is the same at all wavelengths, $(F_{\rm 
n,\lambda}^0+A_{\rm n,\lambda})/(F_{\rm n,\lambda}^0- A_{\rm n,\lambda}) \approx (F_{\rm 
U,0}+A_{\rm U})/(F_{\rm U,0}- A_{\rm U})=C \sim 4.5$, we can separate the contributions of 
the nebular, $F_{\rm n,\lambda}^0$, and the giant, $F_{\rm g,\lambda}^0$, components to  
the constant $F_{\lambda,0}$. Our best fit models are plotted in Fig. 2.

Using the maximum, $F_{\rm \lambda,max} (\phi=0.5)= F_{\rm n,\lambda}^0+A_{\rm 
n,\lambda}$,  fluxes of the nebular component, corrected for the contribution from 
emission lines and the interstellar reddening  the maximum emission 
measure of the nebula, $EM$, has been derived for each band (see M02 for details). These 
$EM$ values,  agree with each other to within a factor of 2, and they cluster around $EM 
\sim 2\times 10^{14}$ and $\sim 3\times 10^{14} {\rm cm}^{-5}$ for $T_{\rm e} = 10^4$ and 
$2\times10^4\,{\rm K}$, respectively. The nebular emission measure derived from our light 
curve analysis requires a Lyman continuum luminosity of $\sim 6$--$5 \times 10^{45}\, 
(d/1\,{\rm kpc})^2\,{\rm phot\,s^{-1}}$ for $T_{\rm e} = 10^4$--$2\times10^4\, {\rm K}$, 
which agrees with independent 
estimates  for the quiescent $L_{\rm h}$ to within a factor $\la 2$ (M02). 

\begin{figure} \plotone{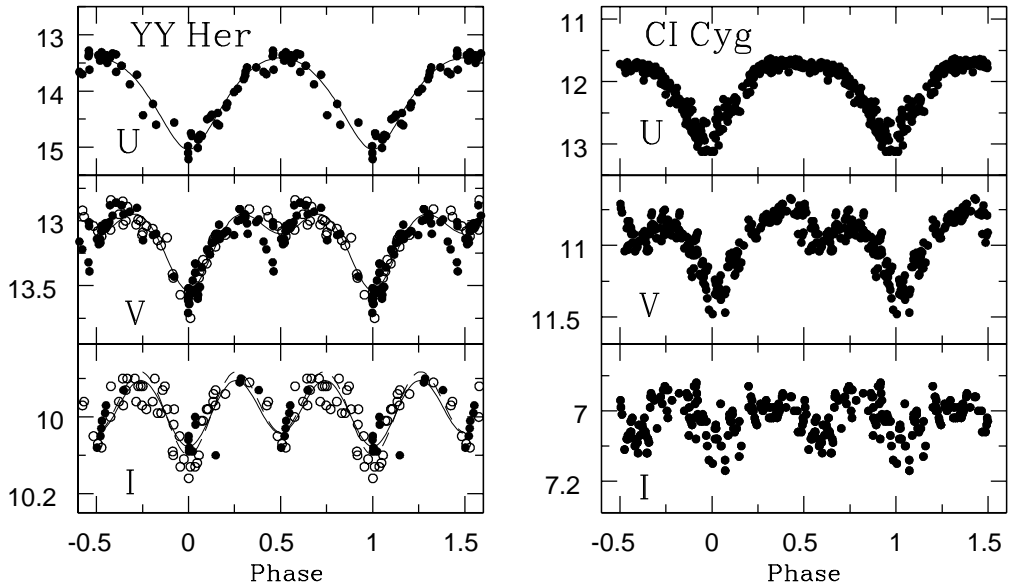} \caption{Phased quiescent  light curves of  CI 
Cyg (Miko{\l}ajewska 2001, and references therein) and YY Her  (points -- M02 data; open 
circles -- H01 data). The ellipsoidal variability is visible only in visual and red light. 
The best fit models (see text) are plotted as solid lines (M02 data) and dashed lines (H01 
data), respectively.} \end{figure}

The  magnitudes, and colours of the cool component derived from our light curve analysis, 
$B-V=1.73$--1.77, $V- R=1.73$--1.92 and $V-I=3.33$--3.61, are in excellent agreement 
with those expected for a moderately reddened M3--M4 giant. Moreover, they indicate 
somewhat higher temperature (earlier spectral type) at maxima ($\phi=0.25$ and 0.75) than 
that at minima ($\phi=0$ and 0.5) as expected for a tidally distorted giant. 

\section{Concluding remarks}

One of fundamental questions in relation of symbiotic binaries is the mechanism that 
powers the Z And-type multiple outburst activity observed in many classical symbiotic 
systems. The quiescent spectra of these systems can mostly be fitted by a hot stellar 
source (most likely a white dwarf powered by thermonuclear burning of the material 
accreted from its companion) and its ionizing effect on a nebula. Their outburst activity, 
however, with time scales of a few/several years cannot be simply accounted by the 
thermonuclear models. A promising interpretation of this activity involves changes in mass 
transfer and/or accretion disc instabilities. Detection of an ellipsoidal hot continuum 
source during outbursts of CI Cyg, AX Per, YY Her, AS 338 and other Z And-type systems 
suggests the presence of an optically thick accretion disc which strongly supports this 
interpretation (Miko{\l}ajewska et al. 2002b, and references therein) 
Related to this problem is the process of mass transfer -- Roche lobe overflow or stellar 
wind -- and possibility of an accretion disc formation. The presence of secondary minima 
in the light curves of YY Her, apparently due to ellipsoidal changes of the red giant, as 
well as in a few other similar systems provides here important clues. Although it is very 
premature to claim that all symbiotic systems with the Z And-type activity do have tidally 
distorted giants, and -- at least during active phase -- 
accretion discs, whereas the non-eruptive systems (such as RW Hya) do not, the former is 
certainly the case for YY Her, CI Cyg, BF Cyg and a few other active systems. 

Systematic searches for the ellipsoidal variations in both active and non-eruptive 
symbiotic stars are necessary to address the problem, and confirm or exclude tidally 
distorted giants. The observations must be, however, carried in the red and near-IR range 
where the cool giant dominated the continuum light. We also note that any optical/red 
light curve analysis must be supported by spectroscopic information about contamination of 
the broad-band photometry by the nebular component. 

\acknowledgments AAT and SYuS would like to thank LOC and SOC for hospitality and financial support.
This research was partly founded by KBN Research Grant No. 
5\,P03D\,019\,20, and RFBR grants Nos. 00-15-96553, 02-02-16462, 
02-02-26635 and 02-02-16235.

\end{document}